# Endogenous Derivation and Forecast of Lifetime PDs

Volodymyr Perederiy*

## Abstract

This paper proposes a simple technical approach for the analytical derivation of Point-in-Time PD (probability of default) forecasts, with minimal data requirements. The inputs required are the current and future Through-the-Cycle PDs of the obligors, their last known default rates, and a measurement of the systematic dependence of the obligors. Technically, the forecasts are made from within a classical asset-based credit portfolio model, with the additional assumption of a simple (first/second order) autoregressive process for the systematic factor. This paper elaborates in detail on the practical issues of implementation, especially on the parametrization alternatives.

We also show how the approach can be naturally extended to low-default portfolios with volatile default rates, using Bayesian methodology. Furthermore, expert judgments on the current macroeconomic state, although not necessary for the forecasts, can be embedded into the model using the Bayesian technique.

The resulting PD forecasts can be used for the derivation of expected lifetime credit losses as required by the newly adopted accounting standard IFRS 9. In doing so, the presented approach is endogenous, as it does not require any exogenous macroeconomic forecasts, which are notoriously unreliable and often subjective. Also, it does not require any dependency modeling between PDs and macroeconomic variables, which often proves to be cumbersome and unstable.

---

* Perederiy Consulting (founder and consultant), PhD (Viadrina University, Germany)
research@perederiy-consulting.de, www.perederiy-consulting.de

# Acknowledgements

I would like to acknowledge the inspirational advice from Jan-Philipp Hoffmann, PhD.

# Introduction and Scope

In 2018, a new accounting standard (International Financial Reporting Standard 9 or IFRS 9) became effective in the EU, setting new rules for accounting of financial instruments (loans, bonds etc.). One of the most important innovations lies in the need to calculate the expected lifetime credit losses (ELCL) for a large class of risky credit exposures (over the entire lifetime of such exposures).

In typical implementation practice, the ELCL for a credit exposure is technically calculated as:

$$ELCL \;\; = \sum_{t=T_0}^{Y} EAD_t \;\; LGD_t \;\; MPD_t \;\; D_t \tag{1}$$

Here, $Y$ stands for the expected lifetime of the exposure in years, and $D_t$ for the discount factor (which is set by the standard to the so-called *effective interest rate* of the credit exposure). The expected lifetime exposure at default $EAD_t$ and the expected lifetime loss given default $LGD_t$ are beyond the scope of this paper. In most cases, they show little stochastics and thus can be calculated using simple assumptions (such as constant LGDs and EADs consistent with the contractual payment schedule of the credit exposure).

The marginal probabilities of default $MPD_t$ represent the major challenge of the ELCL calculation. $MPD_t$ for a future year $t$ is the probability that, of all the possible outcomes, the obligor defaults during this year $t$ (given the information currently available, i.e. at $t = T_0$). These marginal PDs $MPD_t$ can be inferred from the forward PDs $FPD_t$ which reflect the probability that the obligor defaults during the future year $t$, given its survival up to the previous period:

$$MPD_t = FPD_t \, SP_{t-1} \tag{2}$$

with survival probability $SP_{t-1}$ iteratively calculated as:

$$SP_{t-1} = \sum_{j=1}^{t-1} \left(1 - FPD_j\right) \tag{3}$$

using the assumption of an “absorbing” default state (once defaulted, the obligor ceases to exist).

In the modern credit modeling practice, one commonly distinguishes between Point-in-Time (PiT) PDs and Through-the-Cycle (TtC) PDs. The notion of a PiT PD is rather clear: it corresponds to the expected default rate (DR) of an obligor during a specific time period, taking into account all available obligor-specific and macroeconomic information. In contrast, there is some ambiguity as to what the TtC PD actually means (and how it can be estimated or verified using observable data). In this paper, the TtC PD is defined as the unconditional expectation of an obligor default rate, i.e. the expectation in a situation where the macroeconomic conditions are assumed to be completely unknown except for their long-term distribution[1].

According to the IFRS 9 standard, the expectations involved should reflect both obligor-specific and macroeconomic conditions. Thus, the PDs used for the ELCL calculation should clearly be of the PiT type. The forecast of future (forward) PiT PDs is a non-trivial task. Forecasting future macroeconomic conditions themselves is notoriously subjective and unreliable. Moreover, modeling the exact effect of these macroeconomic conditions on PiT PDs is also problematic, as no exact economic theories exist, which makes statistical/empirical modeling necessary. The latter modeling is, however, plagued by various technical challenges, such as trends and non-stationary patterns in data (especially in the case of a short data history), non-consistence of data definitions (especially in the case of a long data history), time lags and leads in relationships, unstable and spurious correlations. This makes the PiT forecasting extremely challenging, and, for long-term forecast horizons, often bordering on crystal-ball clairvoyance.

[1] This interpretation, also used in *Carlehed and Petrov [2012]*, is not the only one possible. For example, *Aquais et al [2008]* interpret the TtC PD as the PiT PD when the non-random macroeconomic state assumes the specific “neutral/normal” level which implies “long-run average default rates”. In our opinion, this latter interpretation is less suitable for the purposes of IFRS 9 PD forecasting, as a future macroeconomic state is uncertain by its nature, which requires the application of statistical expectations to treat it properly.

In contrast, modeling future TtC PDs is far less problematic. The TtC PDs or their approximations are widely available in banking practice. In particular, the external credit ratings and the Basel II internal credit ratings can be regarded as approximate TtC PDs. Standard techniques exist for the projections of the ratings into the future, such as multiplication of rating transition matrices. These established techniques can, in principle, be exploited for the purposes of forecasting. However, especially for long-term forecast horizons, the future TtC of an obligor is also subject to great uncertainty. Hence, not only the expected value, but also the variance of the future TtC PD should be accounted for. If not otherwise indicated, the TtC PDs are assumed to be known in this paper.

## TTC-PIT Transformation: Asset-based Credit Portfolio Approach

A possible theoretical approach to deal with the TtC to PiT transformations draws on the classical asset-based (also referred to as Merton-type) credit portfolio modeling. In this framework, *Vasicek [2002]* derived analytically the asymptotic portfolio loss distribution (also known as ASRF or Asymptotic Single Risk Factor model). In the years following, this asymptotic distribution served as the theoretical underpinnings of Basel II capital requirements for credit risks. On this basis, *Carlehed and Petrov [2012]* investigated in more detail the transformations between TtC , PiT, and intermediate (hybrid) PDs.

The approach is based, in particular, on the concepts of a firm asset return and a default barrier. In this framework, the default of an obligor $i$ in a period $t$ is caused by the value of the obligor's (firm's) assets falling below a certain critical (barrier) level, which is normally linked to the debt of the firm. This condition can be restated as the firm's random asset return $r_{i,t}$ proving to be (in that particular period) lower than a certain critical return $R_{i,t}^{D}$. Thus, the probability of default of the obligor $i$ can be specified as:

$$PD_{i,t} = P\left(r_{i,t} < R_{i,t}^{D}\right) \tag{4}$$

The return $r_{i,t}$ can be assumed to be normally distributed and represented as a weighted sum of a systematic return $\psi_t$ (common to all obligors) and an idiosyncratic return $\varepsilon_{i,t}$ (obligor-specific), with these two return factors being independent of each other. Without loss of generality, after suitable rescaling, the factors can be mathematically expressed as standard normal variables, related as follows:

$$\begin{gathered} r_{i,t} = \psi_t\sqrt{\rho} + \varepsilon_{i,t}\sqrt{1-\rho} \\ \psi_t \sim N(0,1) \\ \varepsilon_{i,t} \sim N(0,1) \end{gathered} \tag{5}$$

Here, the weighting (correlation) coefficient $\rho \in [0;1]$ is the measure of systematic dependence of the obligor. In credit portfolio models, the coefficient is also often referred to as 'R' or 'R-squared'. The unconditional distribution of $r_{i,t}$ (i.e. the distribution when the realization of both $\psi_t$ and $\varepsilon_{i,t}$ is unknown) is also standard normal[2]. The systematic factor $\psi_t$ in the above framework can be interpreted as one single macroeconomic factor affecting all obligors in the portfolio under consideration. Therefore, the Through-the-Cycle (TtC) PD, which is defined above as the expected default rate without the knowledge of the macroeconomic state, amounts for each obligor to the unconditional PD:

$$PD_{i,t}^{TTC} = P\left(r_{i,t} < R_{i,t}^{D}\right) = \Phi\left(R_{i,t}^{D}\right) \tag{6}$$

with $\Phi$ denoting the cumulative standard normal distribution function.

If $PD_{i,t}^{TTC}$ are assumed to be exogenously known (e.g. based on internal or external credit ratings), the implicit unknown barrier $R_{i,t}^{D}$ can be directly inferred from the known TtC PD by inverting the $\Phi$ function:

$$R_{i,t}^{D} = \Phi^{-1}\left(PD_{i,t}^{TTC}\right) \tag{7}$$

On the other hand, the PiT PD assumes the knowledge of the systematic factor $\psi_t$, in particular that it is equal to a certain value $\Psi_t$. Subsequently, this PD can be viewed as the conditional PD:

$$PD_{i,t}^{PIT} = P\left(r_{i,t} < R_{i,t}^{D} \middle| \psi_t = \Psi_t\right) =$$

$$= P\left(\psi_t\sqrt{\rho} + \varepsilon_{i,t}\sqrt{1-\rho} < R_{i,t}^{D} \middle| \psi_t = \Psi_t\right)$$

and, upon substitution from (7) :

$$PD_{i,t}^{PIT} = P\left(\Psi_t\sqrt{\rho} + \varepsilon_{i,t}\sqrt{1-\rho} < \Phi^{-1}\left(PD_{i,t}^{TTC}\right)\right) =$$

$$= P\left(\varepsilon_{i,t} < \frac{\Phi^{-1}\left(PD_{i,t}^{TTC}\right) - \Psi_t\sqrt{\rho}}{\sqrt{1-\rho}}\right)$$

Thus, finally we obtain the transformation equation (for details, see *Carlehed and Petrov [2012]*):

$$PD_{i,t}^{PIT} = \Phi\left(\frac{\Phi^{-1}\left(PD_{i,t}^{TTC}\right) - \Psi_t\sqrt{\rho}}{\sqrt{1-\rho}}\right) \tag{8}$$

## Estimation of Systematic Factor from Default Statistics

The PiT-TtC relationship (8) can be reversed in order to calculate the macroeconomic factor from a known PiT PDs $PD_{i,T_0}^{PIT}$ for a current/previous period $T_0$:

$$\Psi_{T_0} = \left(\frac{\Phi^{-1}\left(PD_{i,T_0}^{TTC}\right) - \Phi^{-1}\left(PD_{i,T_0}^{PIT}\right)\sqrt{1-\rho}}{\sqrt{\rho}}\right) \tag{9}$$

More realistically, the actual individual PiT PDs $PD_{i,T_0}^{PIT}$ of the portfolio obligors would be unknown. The relationship (8), however, can be exploited on a portfolio level, assuming that all obligors in the portfolio have the same systematic risk. In particular, for a portfolio of $N_{T_0}$ obligors, we assume that their TtC one-year PDs $PD_{i,T_0}^{TtC}$ are known for each obligor $i$. We further assume that during the current period $T_0$ a known number of the obligors $N_{T_0}^{D}$ is defaulting. Then, making use of:

---

[2] In particular, $E\left(r_{i,t}\right) = 0 + 0 = 0, Var\left(r_{i,t}\right) = \left(\sqrt{\rho}\right)^2 1 + \left(\sqrt{\rho - 1}\right)^2 1 = 1,$ and it follows: $r_{i,t} \sim N(0,1)$.

$$\mathrm{E}\left(N_{T_0}^{D}\right)=\sum\nolimits_{i=1}^{N_{T_0}} PD_{i,T_0}^{PIT}$$

the unknown systematic factor $\widehat{\Psi}_{T_0}$ can be estimated such that:

$$N_{T_0}^{D} \equiv \sum\nolimits_{i=1}^{N_{T_0}} PD_{i,T_0}^{PIT}=\sum\nolimits_{i=1}^{N_{T_0}} \Phi\left(\frac{\Phi^{-1}\left(PD_{i,T_0}^{TTC}\right)-\widehat{\Psi}_{T_0}\sqrt{\rho}}{\sqrt{1-\rho}}\right) \tag{10}$$

The reliability of the estimates $\widehat{\Psi}_{T_0}$ depends primarily on the magnitude of $\mathrm{E}\left(N_{T_0}^{D}\right)$. For small levels (under 10 or 20), $N_{T_0}^{D}$ might deviate considerably from $\mathrm{E}\left(N_{T_0}^{D}\right)$ in relative $N_{T_0}^{D}/\mathrm{E}\left(N_{T_0}^{D}\right)$ terms, because of the immanent binomial sampling noise. As a consequence, the estimate $\widehat{\Psi}_{T_0}$ from (10) would also be unreliable for low-default portfolios. We will elaborate on this problem later.

## PIT PD: Uncertain Systematic Factor and Forecasts

If the macroeconomic factor $\psi_{T_f}$ in (8) is assumed to be random (stochastic) to some degree, the natural interpretation for the PiT PD would be the statistical expectation:

$$\begin{gathered} PD_{i,t}^{PIT}=E\left(PD_{i,t}^{PIT}\right)= \\ =E\left(\Phi\left(\frac{\Phi^{-1}\left(PD_{i,t}^{TTC}\right)-\psi_t\sqrt{\rho}}{\sqrt{1-\rho}}\right)\right) \end{gathered} \tag{11}$$

The TtC PD $PD_{i,t}^{TTC}$ is assumed here to be known. We further assume a normal distribution for the stochastic factor $\psi_t$ with parameters $E(\psi_t)$ and $Var(\psi_t)$ which reflect, respectively, the expectation and the uncertainty of the macroeconomic conditions. We make the following substitution:

$$x \equiv \frac{\Phi^{-1}\left(PD_{i,t}^{TTC}\right)-\psi_t\sqrt{\rho}}{\sqrt{1-\rho}}$$

Then, $x$ is also normally distributed with parameters as follows:

$$E(x)=\frac{\Phi^{-1}\left(PD_{i,t}^{TTC}\right)-E(\psi_t)\sqrt{\rho}}{\sqrt{1-\rho}}$$

$$Var(x)=\frac{Var(\psi_t)\rho}{1-\rho}$$

Exploiting the following property (see Appendix):

$$\begin{gathered} E\left(\Phi(x)\right)=\Phi\left(\frac{E(x)}{\sqrt{1+Var(x)}}\right) \\ for\ x \sim N\left(E(x),Var(x)\right) \end{gathered} \tag{12}$$

we finally arrive at the following closed-form expression for the PiT PD with a random systematic factor:

$$PD_{i,t}^{PIT}=\Phi\left(\frac{\Phi^{-1}\left(PD_{i,t}^{TTC}\right)-E(\psi_t)\sqrt{\rho}}{\sqrt{1-\rho+Var(\psi_t)\rho}}\right) \tag{13}$$

This expression can, in particular, be used for the prediction of future (forward) PiT PDs for a future period $T_f$, where the future macroeconomic factor $\psi_{T_f}$ would be uncertain by its nature. Assuming a normal distribution for the future uncertain macroeconomic factor $\psi_{T_f}$, with expected (predicted) value $E\left(\psi_{T_f}\right)$ and variance (uncertainty) $Var\left(\psi_{T_f}\right)$, the transformation equation (13) allows us to easily derive the forecasted PiT PD for the period $T_f$. It is important to note that the PiT forecast is affected not only by the expected value of the macroeconomic factor, but also by its variance. In most realistic cases, the nominator $\Phi^{-1}\left(PD_{i,T_f}^{TTC}\right) - E\left(\psi_{T_f}\right)\sqrt{\rho}$ would be negative, and hence, the higher the variance of the future macroeconomic factor, the higher the PiT PD forecast.

Two special cases of the above equation are important. Firstly, an exactly known future macroeconomic factor might be technically described by the assumptions $E\left(\psi_{T_f}\right) = \Psi_{T_f}$ and $Var\left(\psi_{T_f}\right) = 0$. In that case, the expression (13) reduces to the simple conditional PD (as seen in (8)):

$$PD_{i,T_f}^{PIT} = \Phi\left(\frac{\Phi^{-1}\left(PD_{i,T_f}^{TTC}\right) - \Psi_{T_f}\sqrt{\rho}}{\sqrt{1-\rho}}\right)$$

Secondly, if we use the unconditional distribution from the classical setting, i.e. $E\left(\psi_{T_f}\right) = 0$ and $Var\left(\psi_{T_f}\right) = 1$, the expression (13) reduces to the TtC PD forecast:

$$PD_{i,T_f}^{PIT} = PD_{i,T_f}^{TTC}$$

We now return to our factor estimate $\widehat{\Psi}_{T_0}$ in (10) for the current period $T_0$ and assume for now that it is very accurate, so that formally, $\Psi_{T_0}$ can be assumed to be known exactly: $\Psi_{T_0} = \widehat{\Psi}_{T_0}$. For a future forecast period $T_f$, $T_f > T_0$, the exact distribution of $\psi_{T_f}$ conditional on $\Psi_{T_0}$ would generally be unknown, but should conform to the following convergence restrictions:

$$\begin{aligned} for\ \left(T_f - T_0\right) \to 0:\ \ \psi_{T_f}|\Psi_{T_0} \to N\left(\Psi_{T_0}, 0\right) \\ for\ \left(T_f - T_0\right) \to \infty:\ \ \psi_{T_f}|\Psi_{T_0} \to N(0,1) \end{aligned} \qquad (14)$$

If these conditions are satisfied, the systematic factor and the PiT PDs forecasts resulting from (13) would show the continuous pattern seen in practice for default rates and macroeconomic indicators, meaning that for the time shortly after $T_0$, the PiT PD would not differ substantially from the default rate seen in $T_0$, and the macroeconomic conditions can also be assumed to be similar. On the other hand, for a remote forecast period $\left(T_f - T_0 \to \infty\right)$ no assumptions about the macroeconomic conditions can be made, except for their long-term distribution. This latter case results in the PiT PD forecast which equals the TtC PD forecast.

## PIT PD: Autoregressive Systematic Factor and Forecasts

The technical PiT-TtC transformation resulting from the portfolio framework (as specified in equation (8)) is specified independently for each and every period. There are no immediate assumptions regarding serial dependencies and correlations between the systematic factors $\psi_{T1}$ and $\psi_{T2}$ for $T1 \neq T2$.

The intuitively expected continuity for the $\psi_t$ process, along with the convergence criteria for the conditional distribution of $\psi_{T_f}|\Psi_{T_0}$ in (14), can be easily achieved through the assumption of an autoregressive process for the macroeconomic factor $\psi_t$.

*Lamb and Perraudin [2008]* investigated modeling an autoregressive systematic factor in a similar context of capital requirements. In our notations, this study essentially proves that if $\psi_t$ follows an autoregressive process, so will the transformed PiT PD $\Phi^{-1}\left(PD_{i,t}^{PIT}\right)$. The study then shows how the quantiles of the distribution of the PiT PDs can be inferred for the horizon of one year (motivated by the Basel framework). We proceed similarly to *Lamb and Perraudin [2008]*, but instead of focusing on quantiles (which are relevant for the capital requirements), we focus on the expected value of the distribution of future PiT PDs (which is relevant for IFRS 9) and show how it can be calculated analytically. In doing so, our analysis is applicable to arbitrary prediction horizons (motivated by IFRS 9 lifetime considerations).

In its simplest form, the *autoregressive order-1* process (referred to as AR(1) henceforth) is specified for a stochastic variable $x_t$ as follows:

$$x_t = a_0 + a_1 x_{t-1} + \varepsilon_t, \qquad \varepsilon_t \sim N(0, \sigma_\varepsilon) \tag{15}$$

with parameters $a_0$, $a_1$ and $\sigma_\varepsilon$.

The distributional properties of the AR(1) process are as follows[3]. The conditional distribution of $x_{T_f}$ given the knowledge of $x_{T_0} = X_{T_0}$, with $T_f > T_0$ can be shown to be normal with:

$$\begin{gathered} E\left(x_{T_f} \middle| X_{T_0}\right) = X_{T_0} {a_1}^{T_f - T_0} + \frac{a_0}{(1 - a_1)}\left(1 - {a_1}^{T_f - T_0}\right) \\ Var\left(x_{T_f} \middle| X_{T_0}\right) = \frac{1 - {a_1}^{2(T_f - T_0)}}{1 - {a_1}^2} {\sigma_\varepsilon}^2 \end{gathered} \tag{16}$$

Given the stationarity restriction:

$$a_1 \in (-1,1)$$

the unconditional distribution, which is also the asymptotic distribution for $T_f - T_0 \to \infty$, is then described by:

$$\begin{gathered} E\left(x_{T_\infty}\right) = \frac{a_0}{(1 - a_1)} \\ Var\left(x_{T_\infty}\right) = \frac{{\sigma_\varepsilon}^2}{1 - {a_1}^2} \end{gathered} \tag{17}$$

If we now assume the AR(1) specification (15) for the systematic factor $\psi_t$, the convergence criteria (14) can be achieved via suitable parametrization restrictions. In particular, for $T_f - T_0 \to \infty$:

$$\begin{gathered} E\left(\psi_{T_\infty}\right) \equiv 0 \Rightarrow a_0 = 0 \\ Var\left(\psi_{T_\infty}\right) \equiv 1 \Rightarrow {\sigma_\varepsilon}^2 = 1 - {a_1}^2 \end{gathered} \tag{18}$$

Given these restrictions and additionally the restriction:

$$a_1 > 0 \tag{19}$$

the convergence for $T_f - T_0 \to 0$ is also satisfied, as seen from (16). Thus, only one free parameter $a_1$ remains, which also determines the convergence speed. Given the restrictions in (18), the expressions for a conditional distribution in (16), applied to the systematic factor $\psi_{T_f}$, simplify to:

$$\begin{gathered} E\left(\psi_{T_f} \middle| \Psi_{T_0}\right) = \Psi_{T_0} {a_1}^{T_f - T_0} \\ Var\left(\psi_{T_f} \middle| \psi_{T_0}\right) = 1 - {a_1}^{2(T_f - T_0)} \end{gathered} \tag{20}$$

Now, substituting (20) to (13), we finally obtain the forecast for the PiT PD based on the AR(1) assumption for the systematic factor:

[3] See e.g. *Mills [2000], Johnston and DiNardo [1997]*.

$$PD_{i,T_f}^{PIT,AR(1)} = \Phi\left(\frac{\Phi^{-1}\left(PD_{i,T_f}^{TTC}\right) - \Psi_{T_0}\sqrt{\rho {a_1}^{2(T_f-T_0)}}}{\sqrt{1-\rho {a_1}^{2(T_f-T_0)}}}\right) \quad (21)$$

under the parameter restriction:

$$a_1 \in (0,1) \quad (22)$$

Interestingly, the expression in (21) is identical to the simple conditional PiT PD with adjusted (exponentially decaying as $T_f - T_0 \to \infty$) $\rho$ coefficient (compare (21) to (8)).

Another feasible alternative for the process of the systematic factor is *the autoregressive order-2* process (referred to as AR(2) henceforth), which is generally specified for a stochastic variable $x_t$ as follows:

$$x_t = a_0 + a_1 x_{t-1} + a_2 x_{t-2} + \varepsilon_t, \qquad \varepsilon_t \sim N(0, \sigma_\varepsilon) \quad (23)$$

with parameters $a_0$, $a_1$, $a_2$, and $\sigma_\varepsilon$.

The distributional properties of the AR(2) process are as follows[4]. The conditional distribution, given two known observations $X_{T_0}$ and $X_{T_{-1}}$, has no analytical expression, but can be derived iteratively as:

$$\begin{gathered} E\left(x_{T_f} | X_{T_0}, X_{T_{-1}}\right) = a_0 + a_1 E\left(x_{T_f-1} | X_{T_0}, X_{T_{-1}}\right) + a_2 E(x_{T_f-2} | X_{T_0}, X_{T_{-1}}) \\ E\left(x_{T_0} | X_{T_0}, X_{T_{-1}}\right) \equiv X_{T_0} \\ E\left(x_{T_{-1}} | X_{T_0}, X_{T_{-1}}\right) \equiv X_{T_{-1}} \end{gathered} \quad (24)$$

$$\begin{gathered} Var\left(x_{T_f} | X_{T_0}, X_{T_{-1}}\right) = {\sigma_\varepsilon}^2 \sum_{t=1}^{T_f - T_0} {w_t}^2 \\ w_1 = 1, w_2 = a_1 \\ w_t = a_1 w_{t-1} + a_2 w_{t-2} \; for \; t > 2 \end{gathered} \quad (25)$$

Given the (stationarity) parameter restrictions:

$$\begin{gathered} -1 < a_2 < 1 \\ a_2 - a_1 < 1 \\ a_2 + a_1 < 1 \end{gathered} \quad (26)$$

the unconditional distribution, which is also the asymptotic distribution for $T_f - T_0 \to \infty$, can be shown to be normal with parameters:

$$\begin{gathered} E\left(x_{T_\infty}\right) = \frac{a_0}{(1 - a_1 - a_2)} \\ Var\left(x_{T_\infty}\right) = \frac{(1 - a_2)}{(1 + a_2)} \frac{{\sigma_\varepsilon}^2}{(1 - a_2)^2 - {a_1}^2} \end{gathered} \quad (27)$$

Again, in order for systematic factor to satisfy the convergence criteria (14), adopting an AR (2) process requires the following parameter restrictions:

$$\begin{gathered} a_0 = 0 \\ {\sigma_\varepsilon}^2 = \frac{(1 + a_2)((1 - a_2)^2 - {a_1}^2)}{(1 - a_2)} \end{gathered} \quad (28)$$

Given these restrictions and additionally:

[4] See e.g. *Mills [2000], Johnston and DiNardo [1997]*.

$$a_1 > 0 \tag{29}$$

the convergence for $T_f - T_0 \rightarrow 0$ is also satisfied, as can be seen from (23).

The PiT PD forecast based on the AR(2) assumption for the systematic factor then becomes:

$$PD_{i,T_f}^{PIT,AR(2)} = \Phi\left(\frac{\Phi^{-1}\left(PD_{i,T_f}^{TTC}\right) - E_{AR(2)}\left(\psi_{T_f}\middle|\Psi_{T_0},\Psi_{T_{-1}}\right)\sqrt{\rho}}{\sqrt{1-\rho+Var_{AR(2)}\left(\psi_{T_f}\middle|\Psi_{T_0},\Psi_{T_{-1}}\right)\rho}}\right) \tag{30}$$

with the distribution parameters $E_{AR(2)}\left(\psi_{T_f}\middle|\Psi_{T_0},\Psi_{T_{-1}}\right)$ and $Var_{AR(2)}\left(\psi_{T_f}\middle|\Psi_{T_0},\Psi_{T_{-1}}\right)$ iteratively calculated as:

$$\begin{gathered}
E_{AR(2)}\left(\psi_{T_f}|\psi_{T_0,}\psi_{T_{-1}}\right) = a_1 E_{AR(2)}\left(\psi_{T_f-1}|\psi_{T_0,}\psi_{T_{-1}}\right) + a_2 E_{AR(2)}(\psi_{T_f-2}|\psi_{T_0,}X\psi_{T_{-1}}) \\
E_{AR(2)}(\psi_{T_0}|\psi_{T_0,}\psi_{T_{-1}}) \equiv \psi_{T_0} \\
E_{AR(2)}(\psi_{T_{-1}}|\psi_{T_0,}\psi_{T_{-1}}) \equiv \psi_{T_{-1}} \\
Var_{AR(2)}\left(\psi_{T_f}|\psi_{T_0,}\psi_{T_{-1}}\right) = \frac{(1+a_2)((1-a_2)^2 - {a_1}^2)}{(1-a_2)} \sum_{t=1}^{T_f - T_0} {w_t}^2 \\
w_1 = 1, w_2 = a_1 \\
w_t = a_1 w_{t-1} + a_2 w_{t-2}\ for\ t > 2
\end{gathered} \tag{31}$$

and under the following parameter restrictions:

$$\begin{gathered}
-1 < a_2 < 1 \\
a_2 - a_1 < 1 \\
a_2 + a_1 < 1 \\
a_1 > 0
\end{gathered} \tag{32}$$

# TTC PDs: Estimation and Uncertainty

So far, the current and future TtC PDs $PD_{i,T_f}^{TTC}$ have been assumed as known. Although not in the focus of this paper, we will now shorty elaborate on their estimation.

Generally, current TtC PDs can be well approximated from the current external and/or internal credit ratings, which are normally defined as discretized probabilities of default. The external rating agencies (S&P, Moodys, Fitch) explicitly adhere to the TtC approach. TtC is also recommended for the Basel II internal rating models of banks, as the influence of macroeconomic conditions is deemed to be uncertain, volatile, and cyclical. Accordingly, the internal models commonly lack strong macroeconomic inputs and are calibrated to long-term averages of default rates.

The projection of the current TtC PDs into the future can, in principle, be accomplished via the standard method of rating transition matrix exponentiation (matrix powers), see e.g. *Gerhold et al [2017]*. The future (forward) TtC PDs $PD_{i,T_f}^{TTC}$ from this method would show convergence when the forecast horizon $T_f$ increases, where obligors with both currently high and low TtC PDs asymptotically approach some future mid TtC PDs. However, this convergence refers only to the expected value of $PD_{i,T_f}^{TTC}$. For each individual obligor, the future $PD_{i,T_f}^{TTC}$, especially for long forecast horizons, would be subject to great uncertainty. As the relationships (21) and (30) are clearly non-linear in $PD_{i,T_f}^{TTC}$, this uncertainty of a future $PD_{i,T_f}^{TTC}$ should be accounted for when deriving the PiT forecasts. Fortunately, this can easily be achieved using the very methodology of transition matrix exponentiation. In particular, the standard output of this method is also the (multinomial) distribution over non-default rating classes $c$ at an arbitrary future time point, with corresponding class probabilities $P_{i,T_f}^{RAT,c}$. From the rating scale of the relevant rating system, we can deduce the corresponding TtC PDs $PD_c^{TTC}$ for each rating class. Then, the above presented methodology can be accommodated to the uncertain future TtC PDs of obligors via simple weighting:

$$PD_{i,T_f}^{PIT,AR(1)} = \sum_{c=1}^{N_c} \left[ P_{i,T_f}^{RAT,c}\, \Phi\left( \frac{\Phi^{-1}(PD_c^{TTC}) - \Psi_{T_0}\sqrt{\rho {a_1}^{2(T_f - T_0)}}}{\sqrt{1-\rho {a_1}^{2(T_f - T_0)}}} \right) \right] \tag{33}$$

and:

$$PD_{i,T_f}^{PIT,AR(2)} = \sum_{c=1}^{N_c} \left[ P_{i,T_f}^{RAT,c}\, \Phi\left( \frac{\Phi^{-1}(PD_c^{TTC}) - E_{AR(2)}\left(\psi_{T_f} \middle| \Psi_{T_0}, \Psi_{T_{-1}}\right)\sqrt{\rho}}{\sqrt{1-\rho + Var_{AR(2)}\left(\psi_{T_f} \middle| \Psi_{T_0}, \Psi_{T_{-1}}\right)\rho}} \right) \right] \tag{34}$$

with:

$$\sum_{c=1}^{N_c} P_{i,T_f}^{RAT,c} = 1$$

where $N_c$ stands for the overall number of non-default rating classes in the rating system.

That said, the method of transition matrix exponentiation heavily relies on the Markov assumption. According to this assumption, the rating transition probabilities only depend on the current rating, not on previous ratings, and not on the time point. The last stipulated independence might be especially unrealistic for certain credit products, as the dynamics of TtC PDs may also heavily depend on the origination and maturity times. For example, defaults caused by fraudulent behavior typically occur early, and the peak of forward TtC PDs should occur shortly upon origination. In contrast, with bullet loans, the default risk would tend to materialize late, with a peak of forward TtC PDs near the loan maturity. Refinements of the method of transition matrix exponentiation might be needed in such cases.

## Adaptation to Hybrid Ratings

Although most rating systems follow the TtC approach, some show rather PiT or hybrid (falling between TtC and PiT) character. An extreme example would be a rating system which, in short intervals, recalibrates the average portfolio forecasted PD to the observed default rate of the most recent period.

Generally, the 'PiT-ness' can be well integrated into the above framework as follows (see *Carlehed and Petrov[2012]*):

$$PD_{i,t}^{RAT} = \Phi\left( \frac{\Phi^{-1}\left(PD_{i,t}^{TTC}\right) - \Psi_t\sqrt{\rho\lambda^2}}{\sqrt{1-\rho\lambda^2}} \right) \tag{35}$$

with $PD_{i,t}^{RAT}$ being the rating PD and the coefficient $\lambda \in (0,1)$ expressing the PiT-ness of the rating system. The two limiting cases are:

$\lambda$=0 with $PD_{i,t}^{RAT} = PD_{i,t}^{TTC}$

and

$\lambda$=1 with $PD_{i,t}^{RAT} = PD_{i,t}^{PIT}$.

Note that, for a fixed $PD_{i,t}^{TTC}$ and $\Psi_t \sim N(0,1)$, it still holds for an arbitrary $\lambda \in (0,1)$:

$$E\left(PD_{i,t}^{RAT}\right) = PD_{i,t}^{TTC} \qquad (36)$$

consistent with the calibration of the rating systems (incl. hybrid systems) over one or several credit cycles.

It follows from (35):

$$\Phi^{-1}\left(PD_{i,t}^{TTC}\right) = \Phi^{-1}\left(PD_{i,t}^{RAT}\right)\sqrt{1-\rho\lambda^2} + \Psi_t\sqrt{\rho\lambda^2} \qquad (37)$$

Now, after the simple substitution into (8), it holds:

$$PD_{i,t}^{PIT} = \Phi\left(\frac{\Phi^{-1}\left(PD_{i,t}^{RAT}\right)\sqrt{1-\rho\lambda^2} - \Psi_t\sqrt{\rho}(1-\lambda)}{\sqrt{1-\rho}}\right)$$

and, again applying (13), we obtain the PiT forecast with an uncertain systematic factor but a known future rating PD:

$$PD_{i,T_f}^{PIT} = \Phi\left(\frac{\Phi^{-1}\left(PD_{i,T_f}^{RAT}\right)\sqrt{1-\rho\lambda^2} - E\left(\psi_{T_f}\right)\sqrt{\rho}(1-\lambda)}{\sqrt{1-\rho+Var\left(\psi_{T_f}\right)\rho(1-\lambda)^2}}\right)$$

and $PD_{i,T_f}^{PIT,AR(1)}$ and $PD_{i,T_f}^{PIT,AR(2)}$ obtained analogously by imputing corresponding $E\left(\psi_{T_f}\right)$ and $Var\left(\psi_{T_f}\right)$.

More realistically, if the future rating PD is uncertain, the weighting scheme, as specified in (33) or (34), would be faulty, as for a hybrid rating system, the rating class probabilities $P_{i,T_f}^{RAT,c}$ would be correlated with $\psi_{T_f}$ in a complex way. Moreover, the very derivation of $P_{i,T_f}^{RAT,c}$ via usual exponentiation (powers) of rating transition matrices is questionable here (see *Gerhold et al [2017]*). As an extreme example, if the rating system is PiT ($\lambda = 1$) and the $PD_{i,t}^{TTC}$ is constant and the autoregression in the systematic factor is absent ($a_1 = a_2 = 0$), the Markov property is clearly not satisfied for rating transitions: rating deteriorations are very likely to be followed by strong rating improvements (because of, in this case, extreme mean reversion of the systematic factor), and vice versa. This serial dependence is ignored by the matrix exponentiation which typically uses empirical (cohort) rating migration matrices, and would lead, asymptotically, to $P_{i,T_f}^{c,RAT}$ distributions which are too wide, and also to expected future rating PDs which are too high.

Further investigations might be needed to rectify this problem.

## Parametrization Alternatives

The equations (21) and (30)-(31) show how future PiT PDs can be forecasted assuming the AR(1) and AR(2) processes of the systematic parameter respectively, based on the following parameters:

- the current/previous (estimated) macroeconomic risk factors $\Psi_{T_0}$ (for the AR(1) case) or $\Psi_{T_0}$, $\Psi_{T_{-1}}$ (for the AR(2) case)
- the systematic correlation coefficient $\rho$
- autoregressive coefficients $a_1$ (for the AR(1) case) or $a_1$ and $a_2$ (for the AR(2) case), subject to the restrictions in (22) and (32) respectively .

The current systematic factor $\Psi_{T_0}$ and the previous systematic factor $\Psi_{T_{-1}}$ can be e.g. estimated from the known realized default statistics of, respectively, the current and previous periods, as specified in (10). Besides, the estimation of this parameter shall also be revisited in the section on Bayesian estimation below, in the contexts of expert judgments and low-default portfolios. We will now elaborate on how the remaining parameters can be determined in practice.

A natural choice for the correlation coefficient $\rho$ are the values used in asset-based credit portfolio models, either the implicit regulatory portfolio model from the Basel II framework or internal credit portfolio models of banks.

The Basel portfolio model is, in fact, based on theoretical underpinnings which are very similar to the approach presented, and also uses one single systematic factor (see *BCBS [2002]*). The current EU regulations (Basel III as implemented in *EU-CRR [2013])* prescribe the following coefficient values (denoted in the CRR text as 'R' coefficient):

Non-retail obligors (EU-CRR article 153):

- General setting: 12% (for PD=100%) to 24% (for PD=0%)
- Financial large and non-regulated companies: as above, multiplied with a factor of 1.25
- Small/medium companies (SMEs) with sales under EUR 50M: as above, with up to 4% deduction, depending on the firm's sales

Retail obligors (EU-CRR article 154):

- General setting: 3% (for PD=100%) to 16% (for PD=0%)
- Loans secured by real estate (residential mortgages): 15%
- Certain qualifying revolving loans (credit cards and similar): 4%

Hence, the Basel/CRR formulas generally assume some dependence of $\rho$ on the PD (with riskier companies being less dependent on systematic factors). This is quite a questionable assumption, although implementing this dependency would not be problematic in the above framework, as the current and future TtC PDs are assumed to be known. Also, the Basel settings realistically account for the company size: the larger a company, the greater its dependency on macroeconomic conditions and thus the higher the systematic risk.

Also, although the underlying Basel II portfolio model uses has one-year risk horizon, the CRR regulations do account for the loan maturity (see *BCBS [2002]*). For retail exposures, this takes the form of penalizations of the $\rho$ coefficient. This explains why the long-term residential mortgages have such higher settings compared to short-term or cancellable revolving loans. This distorts the CRR 'R' coefficient as a measure for the systematic risk for retail portfolios. In contrast, for non-retail exposures, Basel II accounts for the maturities separately (in the so-called 'M' term), so that the 'R' coefficient captures the systematic risk here seemingly well.

The above Basel II settings for the $\rho$ do have the advantage of an "official" source. However, they were estimated empirically back in the early 2000s using the default statistics collected and aggregated by national banks. For this reason, the contemporary internal credit portfolio models of banks might be a more suitable choice as the source for $\rho$. Most of these models nowadays also follow an asset based approach similar to one presented above, with risk decomposition into systematic and idiosyncratic parts. Many internal asset-based models draw, however, on several correlated systematic factors (not just one single factor as in Basel II), assigning these factors to obligors typically on the basis of their industry and/or region. Nonetheless, if all obligors in the regarded portfolio belong to the same industry and region, the model specifications do not differ much from the single-factor model. The $\rho$ coefficient as assigned to obligors in the internal models also depends on their industry and/or region. Some internal models also account for the company size.

When using the $\rho$ coefficients from the internal model, it is important to ensure that the coefficient is estimated in a way consistent to how it is used for the purposes of TtC-PiT transformations. In particular, data involved for its estimation should cover at least 1-2 credit cycles of default statistics and the PDs used for the estimation should be the TtC PDs. The technical routines for the $\rho$ estimation (typically maximum-likelihood, see e.g. *Kalkbrener & Onwunta [2009]*) do not usually account for the serial correlation patterns. That does not, however, affect the estimates considerably, given sufficiently long estimation data. Generally, for non-retail exposures, the internal models seem to rely on $\rho$ values in the range of 10% to 40%, and for retail exposures in the range of 0.5% to 5%.

Regarding the autoregressive coefficients $a_1$ and $a_2$, there would not be many immediate external sources, and the estimates should rather be based on available data and expert judgements, as well as common knowledge about the macroeconomic credit cycles in the business segments involved. The PiT forecasts from the simple AR(1) process would generally show a mean-reverting behavior, with the coefficient $a_1$ determining the speed of reverting. The lower this coefficient, the sooner the forecasted PiT PDs converge to the corresponding forecasted TtC PDs (in the extreme case, if $a_1 = 0$, all forecasted PiT PDs are equal to the forecasted TtC PDs). The AR(1) process is completely memoryless, in the sense that previous states do not have any influence on the future. In particular, after reverting to its mean, the process is equally likely to move on in either direction, independent of whether the reverting has been from below or above. The AR(1) process also does not show a clear cyclicality/periodicity in the technical sense, as its spectral density does not show a peak. However, the periodicity can be approximated by some proxies. For example, the criterion of “dual-crossing” can be used. With this criterion, the cycle begins when the process crosses its mean from below (or, alternatively, from above) and lasts until the next such crossing from below (or, from above); see Figure 1 in section ‘Illustrations’ below. The mean time lapse between such events can then be regarded as the periodicity proxy.

Using various sources, the range of 0.6 to 0.95 seems to be the most suitable for the AR(1) $a_1$ coefficient. The value of 0.8 results in the dual-crossing periodicity of some 10 years, which also seems to approximate the length of the recent credit cycles (1990-2001, 2001-2009).

Using the more complicated AR(2) process results in losing the simple intuitive closed-from solution for the forward PiT PDs as in the AR(1) case (compare (30)-(31) to (21)). It does, however, offer some advantages. In particular, an AR(2) process can capture the cyclical economic behavior more realistically, because – apart from the mean reverting – it can show the so-called ‘momentum’. When such momentum is present, a process crossing its mean from above (or, alternatively, from below) is more likely to move on below (or, above) the mean before returning to the mean (see Figures 2, 6, 7 vs. Figures 1, 4, 5 in ‘Illustrations’). In technical/statistic terms, the AR(2) process with a suitable parametrization will also show a frequency peak in terms of its spectral density, with the peak frequency calculated as[5]:

$$f = \frac{1}{2\pi} \cos^{-1}\left(\frac{a_1(a_2 - 1)}{4a_2}\right) \qquad (38)$$

The periodicity (in years) is then simply $1/f$. For a periodicity to exist, a further parameter restriction is necessary (which might actually facilitate the parameter choice):

$$a_1{}^2 + 4a_2 < 0 \qquad (39)$$

[5] See *Von Storch and Zwiers [2001].*

The values of $a_1$ in the range from 1.2 to 1.4, and $a_2$ from -0.5 to -0.7 result in a realistic series behavior. With $a_1 = 1.3$ and $a_2 = -0.65$, the theoretical spectral periodicity according to (38) is approximately 10 years (and the dual-crossing periodicity is about 9 years).

## Adaptation to Low Default Portfolios: Bayesian Estimation

The above framework requires the knowledge of the last realized systematic factor $\Psi_{T_0}$ (in the AR(1) case) or the last two factors $\Psi_{T_0}, \Psi_{T_{-1}}$ (in the AR(2) case). In principle, their estimates can be inferred from the known default statistics (as shown in (10) for $\widehat{\Psi}_{T_0}$). The precision of such estimates depends primarily on the number of (expected) defaults in the portfolio. With only a small number (below 10 or 20), the actual/realized number of defaults would be subject to a strong binomial sampling noise. Therefore, the estimates in (10) would be unreliable and volatile in such cases (similar to realized default rates being imprecise estimates of PDs in such cases).

Fortunately, the problem can be efficiently attenuated in the above framework. Instead of using the fixed estimates ($\Psi_{T_0} = \widehat{\Psi}_{T_0}$), we can just assume that $\Psi_{T_0}$ is a random variable with a distribution $f\left(\psi_{T_0}\right)$ which reflects its uncertainty. If we assume $E\left(\psi_{T_0}\right)$ and $Var\left(\psi_{T_0}\right)$ to be the expected value and the variance of this distribution, the distributional properties of the future systematic factors become, for the simple AR(1) case, as follows:

$$
\begin{gathered}
E\left(\psi_{T_f}\right) = E\left(\psi_{T_0}\right) a_1{}^{T_f - T_0} \\
Var\left(\psi_{T_f}\right) = Var\left(\psi_{T_0}\right) a_1{}^{2(T_f - T_0)} + 1 - a_1{}^{2(T_f - T_0)} = \\
= 1 + \left(Var\left(\psi_{T_0}\right) - 1\right) a_1{}^{2(T_f - T_0)}
\end{gathered}
\qquad (40)
$$

Note that, compared to the formulae with a fixed current estimate $\Psi_{T_0}$ (20), the above adjustments:

- substitute the fixed estimate with its expected value for a short-term expectation
- do not change the long-term expectation (which still converges to 0)
- inflate the short-term variance of the systematic factor
- do not change the long-term variance (which still converges to 1).

If $f\left(\psi_{T_0}\right)$ is normal or approximately normal, so will be the distribution of $\psi_{T_f}$, and the expected value and variance of $\psi_{T_f}$ can be directly used in (13) to arrive at the PiT PD forecasts, which now account for the uncertainty in the current systematic factor $\Psi_{T_0}$.

Now, turning our attention to the distribution $f\left(\psi_{T_0}\right)$ itself, the natural choice might be Bayesian-type modeling. In the Bayesian framework, there is an initial belief (in form of a 'prior' distribution) about a parameter to estimate. That belief changes (to a 'posterior' distribution) if new evidence is encountered. We can assign (omitting the $T_0$ index):

- the uncertain current systematic factor $\Psi$ as the parameter to estimate
- the unconditional distribution $N(0,1)$ as its prior distribution
- the default count $N^D$ (out of $N$ obligors) as data evidence
- the distribution $f\left(\psi | N^D\right)$ as the posterior distribution of the systematic factor

We also assume the knowledge of TtC PDs (in this case, common for all obligors) and $\rho$ coefficient.

Then, the posterior distribution can be estimated using the Bayes rule:

where $B$ stands for the density of binomial distribution.

$$f(\psi|N^D) = \frac{f(\psi)f(N^D|\psi)}{f(N^D)} \tag{41}$$

with:

$$\begin{gathered} f(N^D) = \int f(N^D|\psi)f(\psi)d\psi \\ f(N^D|\psi) = B(N, N^D, PD^{PIT}|\psi) \\ PD^{PIT}|\psi = \Phi\left(\frac{\Phi^{-1}(PD^{TTC}) - \psi\sqrt{\rho}}{\sqrt{1-\rho}}\right) \end{gathered} \tag{42}$$

In all likelihood, the posterior distribution seems to be quite close to normal/Gaussian (see Figure 3). Upon calculation, its expected value and variance of this distribution can be estimated and used for the Bayesian forecast of PiT PDs as outlined in (40) above.

It is important to stress that the Bayesian approach inherently accounts for the sample size. Holding the default rate $N^D/N$ constant, an increasing $N$ leads to a narrower posterior distribution with its peak getting more extreme and approaching the "realized" systematic factor Ψ (which is implied by $PD^{PIT}|\Psi \equiv N^D/N$), see Figure 3 in 'Illustrations'. Generally, the Bayesian approach results in PiT forecasts which are more consistent with the current unobserved PiT PDs than the simple forecasts derived from the observed default rates according to (10) (see Figures 4 and 5 in 'Illustrations').

Lastly, the above Bayesian methodology allows for an easy way to embed expert judgements about the current macroeconomic environment. To achieve this, the prior *N(0, 1)* simply needs to be appropriately adjusted, e.g. to *N(-1, 0.5)* for an expert judgement of a mild economic depression with a light uncertainty.

## Compliance of the Approach with Legal Regulations

The presented approach seems to comply well with the detailed legal requirements in the IFRS 9 standard (*EU-IFRS [2016])* and subordinate legislation[6].

In particular, the standard requires that (see section 5.5.17 of the standard):

*An entity shall measure expected credit losses of a financial instrument in a way that reflects:*

(a) *an unbiased and probability-weighted amount that is determined by evaluating a range of possible outcomes*
(b) *the time value of money; and*
(c) *reasonable and supportable information that is available without undue cost or effort at the reporting date about past events, current conditions and forecasts of future economic conditions.*

IFRS 9 further defines (B5.5.49-51) the reasonable and supportable information as "*reasonably available at the reporting date without undue cost or effort, including information about past events, current conditions and forecasts of future economic conditions*", with no need to "*undertake an exhaustive search*".

The probability weighting of macroeconomic outcomes is achieved through the assumption of an autoregressive process for the systematic factor, which results in the future factors being a stochastic variable covering all possible macroeconomic outcomes. The past events are implicitly accounted for in the estimation of the current/past factors $\Psi_{T_0}, \Psi_{T_{-1}}$. The forecasts are thus basically a technical extrapolation of these (under autoregressive model parametrization), which is a reasonable approach given that exact macroeconomic forecasts normally show (at best) only very slight advantages compared to technical or even naïve forecasts (such as the last known conditions). The standard even explicitly states (B5.5.52) that "*in some cases, the best reasonable and supportable information could be the unadjusted historical information*" and (B5.5.50) "*the estimate … does not require a detailed estimate for periods that are far in the future-for such periods, an entity may extrapolate projections form available, detailed information*."

[6] See *BIS [2015], GPPC [2016], IFRS-Project [2015], EBA [2017].*

The standard also reads (B5.5.51) that "*The information used shall include factors that are specific to the borrower, general economic conditions and an assessment of both the current as well as the forecast direction of conditions…"*. This also seems to correspond well to the presented approach, as it draws on TtC PDs, macroeconomic factors, and autoregressive extrapolation respectively.

Finally, the standard states (B5.5.51) that "*Possible data sources include internal historical credit loss experience, internal ratings...".* This is also captured through embedding the default statistics and the Basel-II ratings in the presented approach.

## Summary and Outlook

This paper proposed a simple technical approach for the derivation of future (forward) Point-in-Time PD forecasts. The approach relies on the classical credit portfolio framework (using a single macroeconomic factor) and can therefore draw on a parameter set available from portfolio modeling. The approach uses an autoregressive specification for the systematic factor. In the case of simple AR(1) specification, the resulting PiT PDs are particularly intuitive and correspond to the classical conditional Vasicek PD, with the correlation coefficient exponentially decaying in time. The more advanced AR(2) specification allows for embedding of momentum and clear periodicity into the dynamics of the systematic factor. The approach remains purely analytical and does not require any Monte Carlo simulations.

The resulting Point-in-Time PD forecasts incorporate the expected macroeconomic trends and account for their typically mean-reverting and/or cyclical nature. In doing so, the approach implicitly factors in the great uncertainty involved in forecasting macroeconomic dynamics. The approach works without having to explicitly model the linkage between PDs and various macroeconomic factors, which often proves to be cumbersome and unstable.

The resulting Point-in-Time PDs can be used in many credit-risk related fields, and can also be easily integrated into the calculation of Lifetime Expected Credit Losses as required by the new IFRS 9 accounting standard. The approach can also be adopted to low-defaulted portfolios.

The approach assumes the knowledge of (current and future) Through-the-Cycle PDs (or their distribution), typically inferred from credit ratings and rating transition matrices. After minor adjustments, the approach remains applicable even in the case of hybrid-type ratings (falling between Point-in-Time and Through-the-Cycle definitions). More rigorous work, however, is needed to elaborate in detail on the optimal capturing of the TtC PD uncertainty/predictions and hybrid ratings.

# Illustrations

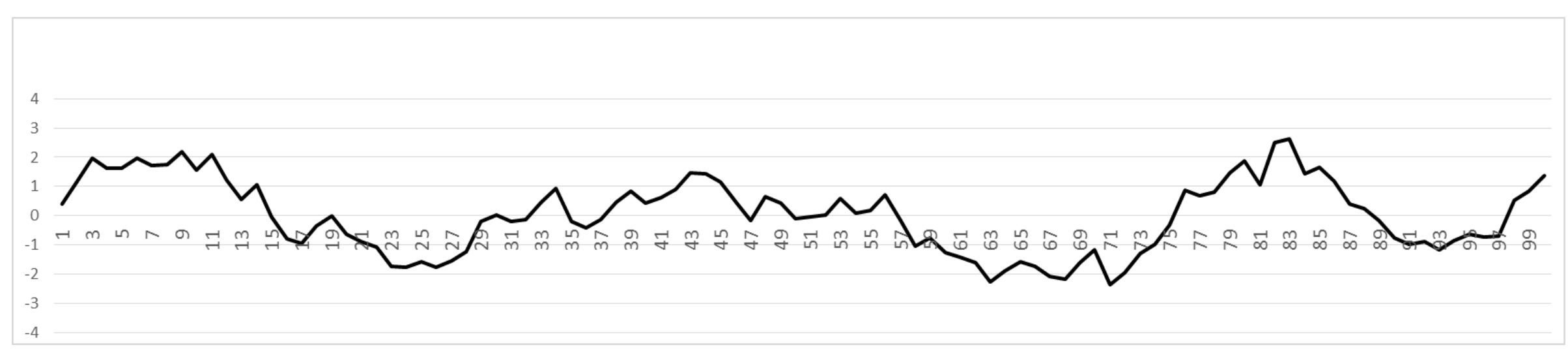

*Figure 1: Simulated realizations for 100 years of systematic factor following AR(1) process with a1=0.8. Using the criterion of "dual crossing", a full "cycle" begins e.g. in the year 75 and ends in the year 97.*

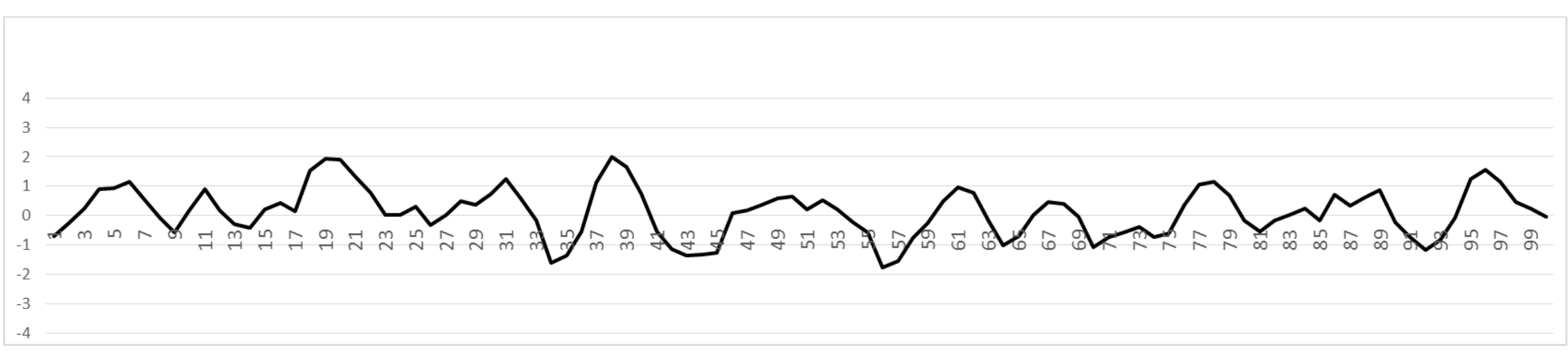

*Figure 2: Simulated realizations for 100 years of systematic factor following AR(2) process with a1=1.3 and a2=-0.65.*

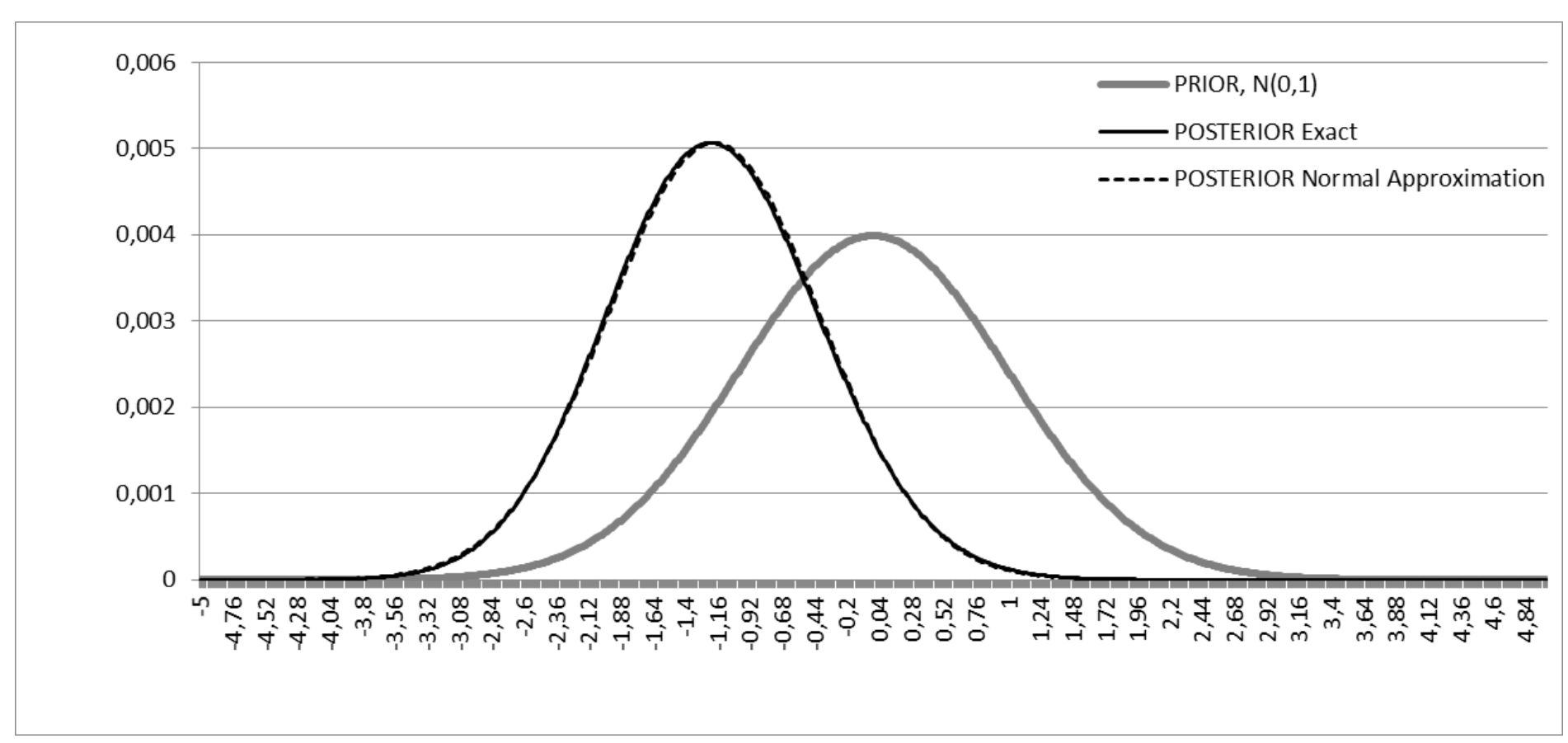

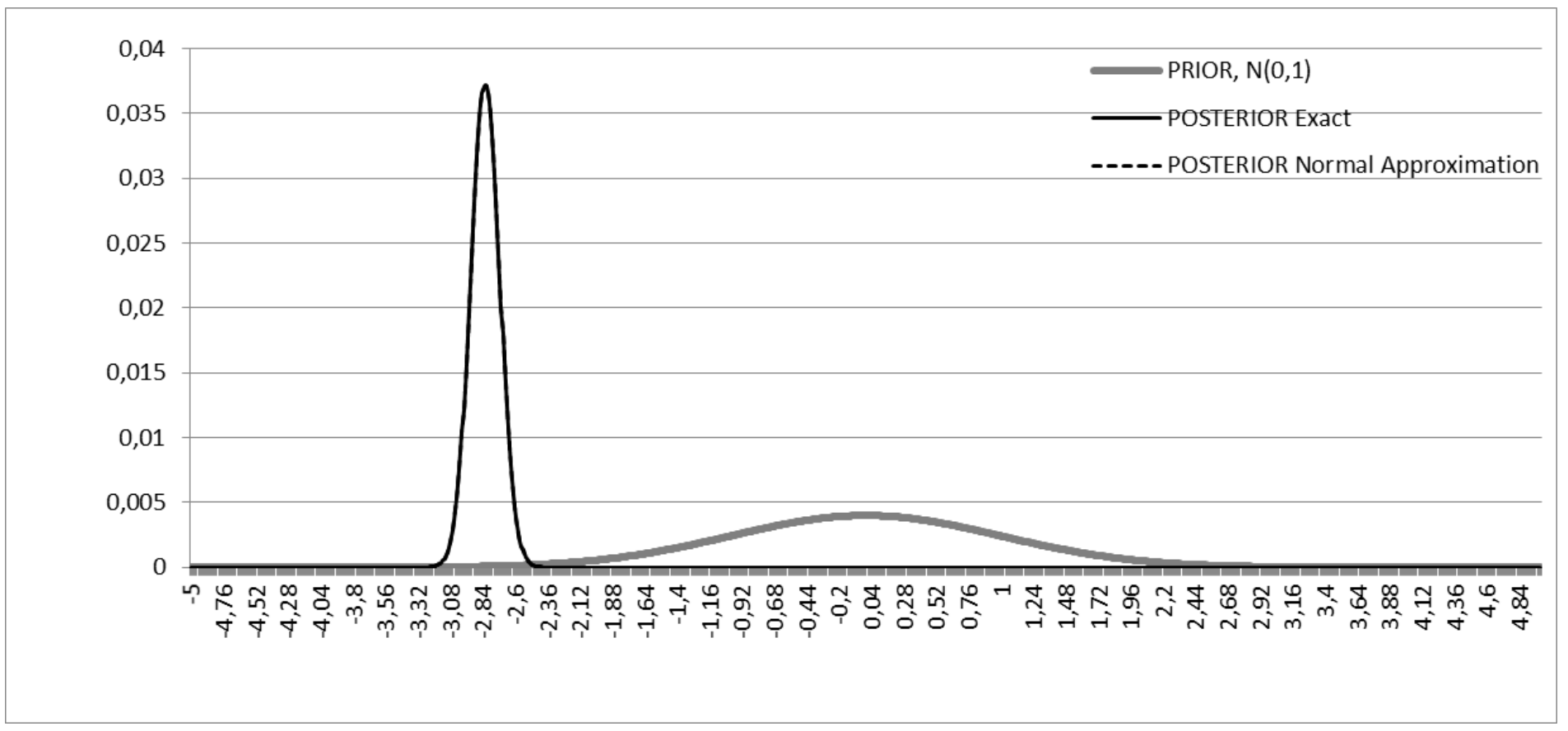

*Figure 3: Bayesian prior, posterior exact, and posterior approximated distributions for the systematic factor, with $PD^{TTC} = 3\%, \rho = 15\%, N = 10\ (above)\ or\ N = 1000\ (below)$, and observed default rate $(N^D/N)$ of 20%.*

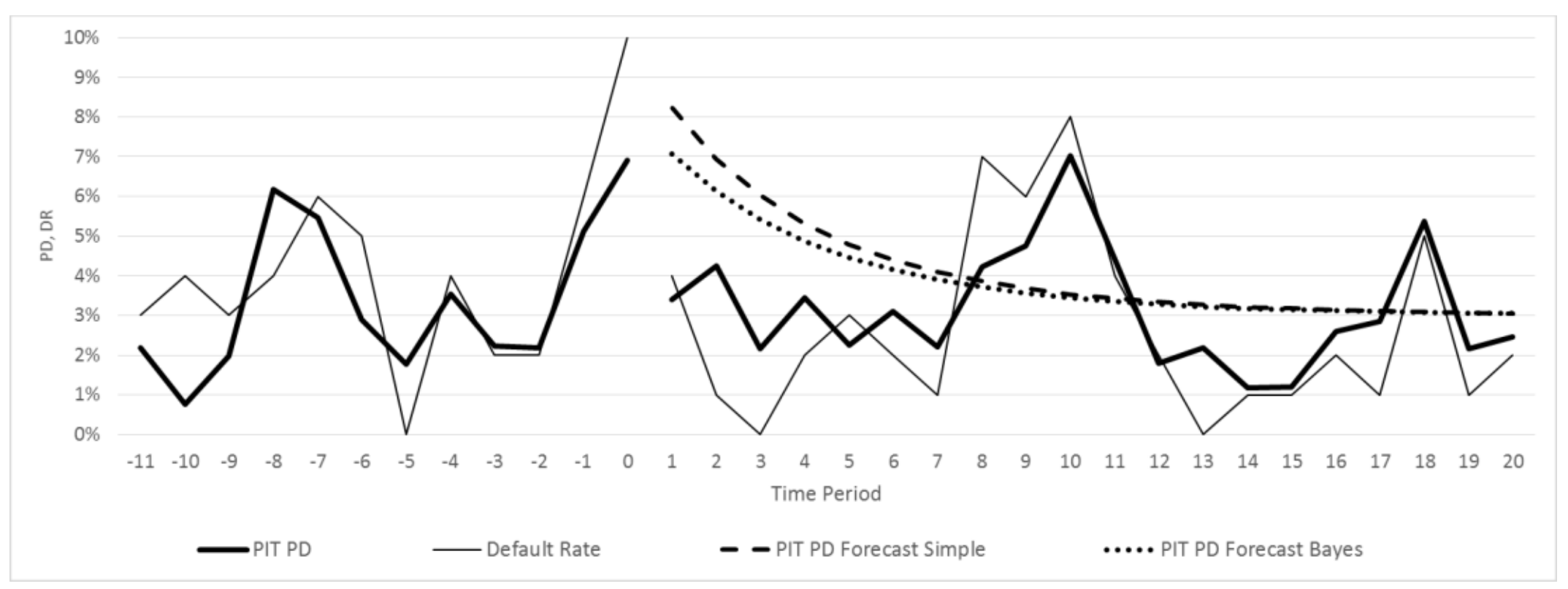

*Figure 4: Simulated PiT PDs and default rates, as well as forecasted (simple and Bayes) PiT PDs (as of T=0) for:* $PD^{TTC} = 3\%, \rho = 15\%, N = 100$*, AR(1) process with a1=0.8.*

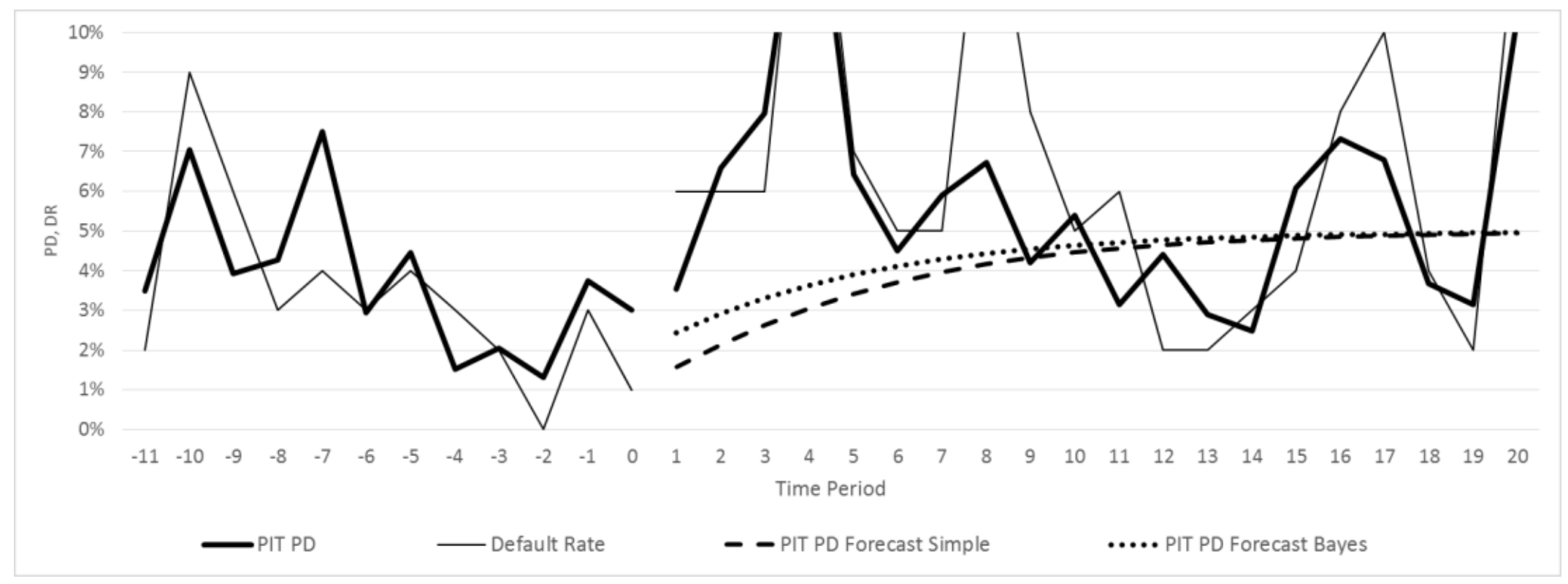

*Figure 5: Simulated PiT PDs and default rates, as well as forecasted (simple and Bayes) PiT PDs (as of T=0) for:* $PD^{TTC} = 5\%, \rho = 15\%, N = 100$*, AR(1) process with a1=0.8.*

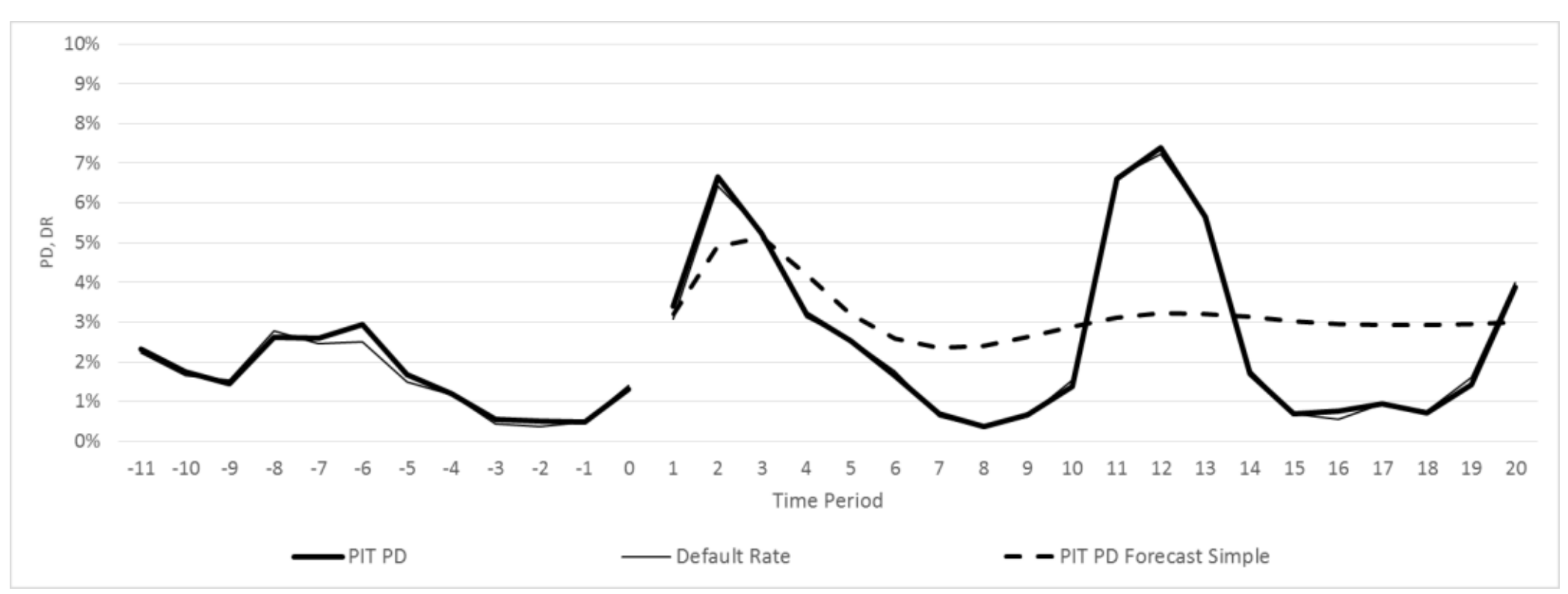

*Figure 6: Simulated PiT PDs and default rates, as well as forecasted PiT PDs (as of T=0) for:* $PD^{TTC} = 3\%, \rho = 15\%, N = 10{,}000$*, AR(2) process with a1=1.3 and a2=-0.65.*

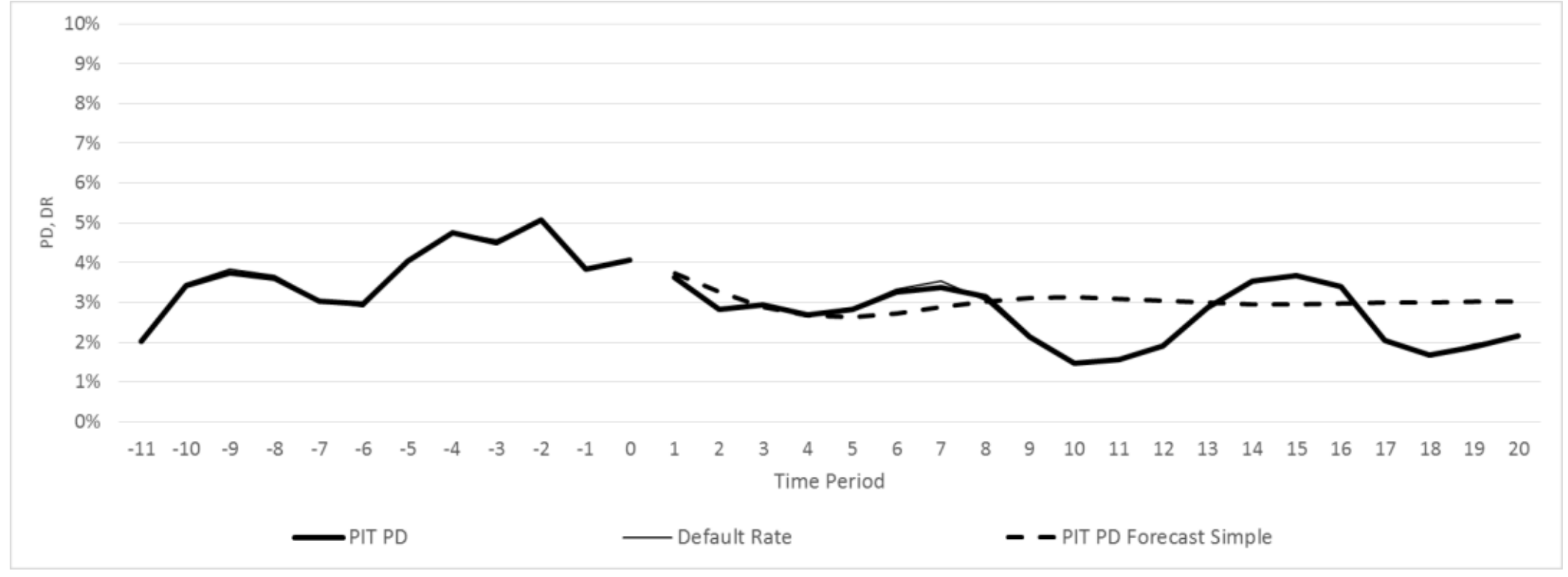

*Figure 7: Simulated PiT PDs and default rates, as well as forecasted PiT PDs (as of T=0) for:* $PD^{TTC} = 3\%, \rho = 3\%, N = 100{,}000$*, AR(2) process with a1=1.3 and a2=-0.65.*

# Appendix: Expected Value of Normal Cumulative Function

For $x \sim N(\mu, \sigma)$:

$$E\big(\Phi(x)\big) = \Phi\left(\frac{\mu}{\sqrt{1+\sigma^2}}\right)$$

**Proof:**

$$E\big(\Phi(x)\big) = E\big(\Phi(\mu + \sigma y)\big), \qquad y \sim N(0,1)$$

By definition of $\Phi$, for an independent variable $z \sim N(0,1)$, it holds:

$$E\big(\Phi(x)\big) = E\big(P(z \le \mu + \sigma y)\big)$$

Then:

$$E\big(\Phi(x)\big) = E\big(P(z - \sigma y \le \mu)\big) = P(z - \sigma y \le \mu)$$

$z - \sigma y$ is normally distributed with $E(z - \sigma y) = 0$ and $Var(z - \sigma y) = 1 + \sigma^2$.

Therefore, an expression $\frac{z - \sigma y}{\sqrt{1+\sigma^2}}$ has the standard normal distribution.

Finally, we obtain:

$$E\big(\Phi(x)\big) = P\left(\frac{z - \sigma y}{\sqrt{1+\sigma^2}} \le \frac{\mu}{\sqrt{1+\sigma^2}}\right) =$$

$$= \Phi\left(\frac{\mu}{\sqrt{1+\sigma^2}}\right)$$